\documentclass[preprint2,usenatbib]{aastex}

\def\mmm{(m-M)$_0$~}
\def\ebv{E(B$-$V)~}

\def\msun{M$_{\odot}$}
\def\gsim{\;\lower.6ex\hbox{$\sim$}\kern-7.75pt\raise.65ex\hbox{$>$}\;}
\def\lsim{\;\lower.6ex\hbox{$\sim$}\kern-7.75pt\raise.65ex\hbox{$<$}\;}

\shorttitle{The BOCCE project}
\shortauthors{Bragaglia \& Tosi}

\received{2005 September 11}
\begin{document}

\title{The Bologna Open Cluster Chemical Evolution (BOCCE)
Project: 
\\ midterm results from the  photometric sample}

\author{Angela Bragaglia  \& Monica Tosi}
\affil{INAF-Osservatorio Astronomico di Bologna, via Ranzani 1, I-40127 Bologna,
      Italy}
\email{angela.bragaglia@bo.astro.it,  monica.tosi@bo.astro.it}
\date{}

\begin{abstract}
We describe a long term project aimed at deriving information on the chemical
evolution of the Galactic disk from a large sample of open clusters. The main
property of this project is that all clusters are analyzed in a  homogeneous
way, to guarantee the robustness of the ranking in age, distance  and
metallicity. Special emphasis is devoted to the evolution of the earliest 
phases of the Galactic disk evolution, where clusters have superior
reliability with respect to other types of evolution indicators. The project
is twofold: on the one hand we derive age, distance and reddening (and
indicative metallicity) interpreting deep and accurate photometric data with
stellar evolution models, and, on the other hand, we derive the chemical
abundances from high-resolution spectroscopy. Here we describe our overall
goals and approaches, and report  on the mid-term project status of the
photometric part, with 16 clusters already studied, covering an age interval
from 0.1 to 6 Gyr and  Galactocentric distances from 6.6 to 21 kpc. The
importance of quantifying the theoretical uncertainties by deriving the
cluster parameters with various sets  of stellar models is emphasized. Stellar
evolution models assuming overshooting from convective regions appear to
better reproduce the photometric properties of the cluster stars.

The examined clusters show a clear metallicity dependence on the
Galactocentric distance and no dependence on age. The tight relation between
cluster age and magnitude difference between the main sequence turn off and
the red clump is confirmed. 

\end{abstract}


\section{Introduction}
In the last decade, our understanding of the chemical evolution of the Galaxy
has significantly  improved, thanks to the efforts and the  achievements both
on the  observational and on the theoretical sides. Numerical chemical 
evolution models nowadays can satisfactorily reproduce the major  observed
features of the Milky Way (e.g.  Lacey \& Fall 1985, Tosi 1988,  Matteucci \&
Fran\c cois 1989, Ferrini et al. 1994, Giovagnoli \& Tosi 1995, Timmes,
Woosley \& Weaver 1995, Chiappini, Matteucci \& Gratton 1997, Boissier \&
Prantzos 1999).  They are  considered successful only when they reproduce 
the  current distribution with Galactocentric distance of the SFR (e.g. as
compiled by Lacey \& Fall 1985), the current distribution with Galactocentric
distance of the gas and star      densities  (see e.g. Tosi 1996, Boissier \&
Prantzos 1999 and references  therein); the current distribution with
Galactocentric distance of element abundances as derived from HII regions and
from B-stars (e.g. Shaver et al. 1983, Smartt \& Rolleston 1997); the
distribution with Galactocentric distance of element abundances at slightly
older epochs, as derived from PNe II (e.g. Pasquali \& Perinotto 1993, Maciel
\& Chiappini 1994, Maciel \& K\"oppen 1994, Maciel, Costa \& Uchida al. 2003);
the  age-metallicity relation (AMR) not only in the solar neighborhood (Twarog
1980) but also at other distances from the Galactic center (e.g. Edvardsson et
al. 1993); the metallicity distribution of G-dwarfs in the solar neighborhood
(e.g. Rocha-Pinto \& Maciel 1996); the local Present-Day-Mass-Function (PDMF,
e.g. Scalo 1986, Kroupa, Tout \& Gilmore 1993); the relative abundance ratios
(e.g. [O/Fe] vs [Fe/H]) in disk and halo stars (e.g. Barbuy 1988, Edvardsson
et al. 1993).

There are, however, several open questions which still need to be answered.
One of the important longstanding questions concerns the evolution of the
chemical abundance gradients in the Galactic disk. The distribution of heavy
elements with Galactocentric distance, as derived from young objects like HII
regions or B stars, shows a  trend in the disk of the Galaxy and of other well
studied spirals which is usually interpreted as a linear negative gradient.
It  is not clear yet whether the gradient slope changes with space and time,
and, in  case, if it flattens or steepens.

Galactic chemical evolution models do not provide a consistent answer: even
those that are able to reproduce equally well the largest set of observational
constraints predict different scenarios for early epochs. By comparing with
each other the most representative models of the time, Tosi (1996) showed that
the predictions on the gradient evolution range from a slope initially
positive which then becomes negative and steepens with time, to a slope
initially negative which remains roughly constant, to a  slope initially
negative and steep which gradually flattens with time (see also Tosi 2000 and
Chiappini, Matteucci \& Romano 2001 for updated discussions  and references).

From the observational point of view, the situation is not much clearer.  Data
on field stars are inconclusive, due to the large uncertainties  affecting the
older, metal poorer ones. Planetary nebulae (PNe) are better  indicators,
thanks to their brightness. PNe of type II, whose progenitors  are on average 
2--3 Gyr old, provide information on the Galactic  gradient at that epoch and
show gradients similar to those  derived from HII regions. However, the
precise slope of the radial abundance distribution, and therefore its possible
variation with time is still subject of debate. The PNe data of Pasquali \&
Perinotto (1993),  Maciel \& Chiappini (1994) and Maciel \& K\"oppen (1994)
showed gradient slopes slightly flatter than those derived from HII regions,
but Maciel et al. (2003) and Maciel, Lago \& Costa (2005) have recently
inferred, from a larger and updated PN data set, a flattening of the oxygen
gradient slope during the last 5--9 Gyr.

We believe that open clusters (OCs) represent the best tool to understand 
whether and how the metallicity spatial distribution changes with time, since 
they have formed at all epochs and their ages, distances and metallicities
are  more safely derivable than for field stars.  OCs have generally been
suggested to show a metallicity gradient  (e.g., Janes 1979, Panagia \& Tosi
1981, Friel  1995, Carraro, Ng \& Portinari  1998, Friel et al. 2002), and
Maciel et al. (2005) have recently suggested that OC literature data are
consistent with those on PNe  in the indication of a time flattening of the
gradient.  However, the spatial and temporal evolutions of the metallicity
distribution of OCs are still uncertain.

Bragaglia et al. (2000) using the compilation of ages, distances and 
metallicities listed by Friel (1995), and dividing her clusters in four age 
bins, found no significant variation in time of the gradient slope. This,
however, more than reflecting the Galaxy evolution, may be an effect of 
inhomogeneity in  the data treatment of clusters taken from different
literature sources. Carraro et al. (1998), using the metallicities of 37 OCs 
derived by Friel's group from low resolution spectra, and age and distances
based on their own synthetic color-magnitude diagrams (CMDs), tried to
measure the time  dependence of the radial metallicity gradient. They found
hints that the gradient may be slightly shallower at the present time than in
the past, but, again, the results may be affected by uncertainties related to
the uneven  quality of the literature data they applied the synthetic CMD
method to. Similarly, Friel et al. (2002) presented new, improved data on the
metallicity of 39 OCs, and derived a smooth radial gradient, with indication
that its slope may be steeper for old clusters, but do not consider this a
firm result, due to the different space distribution of young and old OCs. 

Against the  view of an OC's radial abundance gradient,   Twarog, Ashman, \&
Anthony-Twarog (1997), and more recently Corder \& Twarog (2001), argue that
what is observed is a step distribution in OC metallicity, rather than a
smooth gradient: clusters with R$_{GC} \le$ 10 kpc have solar metallicity
(with some dispersion), while clusters farther away have [Fe/H] $\sim -0.3$
dex (again with some dispersion). Yong, Carney, \& Teixera de Almeida (2005)
also find a discontinuity at 10 kpc in the radial distribution of the OC
metallicity, and a {\it basement} [Fe/H]$\simeq$-0.5 beyond that radius, but
agree on the existence of a smooth gradient in the inner regions.

Open clusters are also good tracers of the Galaxy formation mechanism (Freeman
\& Bland-Hawthorn 2002), in particular of the thin disk.  The old OCs may date
the Galactic disk, as an alternative to White Dwarfs, which still give a wide
range of values, depending  on the adopted cooling models (Hansen \& Liebert
2003).  OCs do not offer only a value integrated on the whole disk, but give
information on its different parts. If the discontinuity at 10--12 kpc
described above is real, it may represent the limit of the original thick disk
or the connection of OCs  to the presence of accreted satellites (Frinchaboy
et al. 2004, 2005;  Bellazzini et al. 2004)  or to the occurrence of major
merger events (Yong et al. 2005).  These possibilities are not mutually
excluding, since the accretion-related clusters would be those in the outer
part of the Galaxy, where the concept of "true" disk is fuzzy. As already
noted by Friel (1995), old OCs may be the few survivors of the original ones
that formed in the early Galaxy, or may be the product of later accretion or
interaction with merging satellites.

 Finally, OCs are excellent tests for stellar evolution theory, since they
contain simple stellar populations (i.e. coeval stars with equal metallicity).
Viable stellar evolution models must therefore reproduce  the various features
(morphology, number counts, color and luminosity functions) of their CMDs.
Comparing theoretical and observational CMDs one can discriminate among
different models and select the most realistic ones.  The method of using
synthetic CMDs to compare with observational  data has proven to be much more
informative and rewarding  (Aparicio et al. 1990, Tosi et al. 1991, Skillman \&
Gallart 2002) than the classical isochrone fitting.   These MonteCarlo
simulations allow modeling of  several additional parameters which dictate
the distribution of points  in the CMD, such as stochastic star formation (SF)
processes, binary fraction,  photometric spread, main-sequence (MS) thickness,
data incompleteness and small  number statistics.  Consequently, the results
not only provide a measure of  the properties of the cluster, but can also
constrain the star formation  history (SFH) and the initial mass function
(IMF).  

To try to improve our understanding in all these subjects it is  crucial to
build up a large sample of open clusters for which the age, distance and
metallicity are all derived in the most precise and homogeneous way, to avoid
spurious effects due to inhomogeneous treatments.  For this reason, we are
undertaking a long term project of accurate photometry and high-resolution
spectroscopy to derive homogeneously ages,  distances, reddening and element
abundances in open clusters of various ages and Galactic locations, and
eventually infer from them the  evolution of the metallicity radial
distribution. To this aim we need to collect at least 30 clusters, evenly
distributed in age, metallicity and Galactocentric distance bins to allow for
reliable derivation of possible trends (or lack thereof). Age, distance, and
reddening  are obtained  deriving the CMDs from deep, accurate photometry and
applying to them the synthetic  CMD method described by Tosi et al. (1991).
Accurate chemical abundances are obtained from high-resolution spectroscopy,
applying to all clusters the same method of analysis and the same metallicity
scale. 

Here we describe the results obtained for the first half of the photometric
sample. Spectroscopic abundances have been presented so far only for  NGC 
6819 (Bragaglia et al. 2001), NGC  2506, NGC  6134, IC 4651 (Carretta et al.
2004), and Cr 261 (Carretta et al. 2005) and will be the subject of
fore-coming papers. This paper is organized as follows: the BOCCE photometric
sample in described in Section 2 and the  application of the synthetic CMD
method in Section 3.  The results obtained so far by BOCCE on age, distance
and reddening of  the examined clusters are summarized in Section 4, and
discussed in Section 5 in terms of test for the stellar evolution models.
Contaminating effects are examined in Section 6.  A discussion of the mid-term
evolution picture obtained from the currently assembled sample is provided in
Section 7 and a summary in Section 8.

\section{The photometric sample}  

Homogeneity is a fundamental requirement when using different data sets to
make  detailed comparisons between observational data and models. To truly
understand the variations in the different model parameters, one must minimize
systematic errors. What is needed for the purposes described above is a large
data set of open clusters, all observed with the same technique, analyzed with
the same method, accurately  and consistently calibrated in the same
photometric system, and with well determined incompleteness factors and
photometric errors. Our photometric  sample is being built following these
requirements.

Up to now we have acquired the photometry for more than 30 clusters, in some
cases only in the B and V bands, in others also in the I band, and in a few
cases also in U and additional bands. About half have been analyzed.
All magnitudes have been calibrated to
the Johnson-Cousins system, via observations of primary and secondary standard
stars (Landolt 1992, Stetson 2000) properly selected to cover the color range
of the cluster stars. All the data have been reduced with PSF fitting
packages, particularly  suitable for crowded fields photometry; we have used
DAOPHOT (Stetson 1987, Davis 1994) in most cases, ROMAFOT  (Buonanno et al.
1983) and PSFex (E. Bertin, priv. communication) in a few others. In all cases
incompleteness factors and photometric errors have been determined with the
help of extensive artificial star tests. All the details can be found in the
original papers quoted in Table 1.

We have already  published the results for 16 of these clusters\footnote {All
the published data are available in electronic form from the BDA (Base Des
Amas) maintained by J.C. Mermilliod at
http://obswww.unige.ch/webda/webda.html}, which are listed in Table 1,
together with the derived age, distance modulus, reddening, indicative
metallicity and reference. For sake of homogeneity, all the listed quantities
refer to the best cases obtained with synthetic CMDs (see next Sections) based
on the Padova 1994  stellar models (Bressan et al. 1993, Fagotto et al. 1994),
because these are the tracks which more often turned out to provide the best
fit to the data. There are (a few) cases,  however, where other stellar models
were found to better reproduce the data (see next Sections and the papers
quoted in  Table 1 for details).

The metallicity indicated in Table 1 is that of the stellar models in better
agreement with the data and cannot be taken as an accurate determination of
the cluster elemental abundance, for which detailed spectroscopic analysis is
necessary. 

\begin{figure}
\includegraphics[bb=25 190 515 670, clip=true, angle=0, scale=.95]{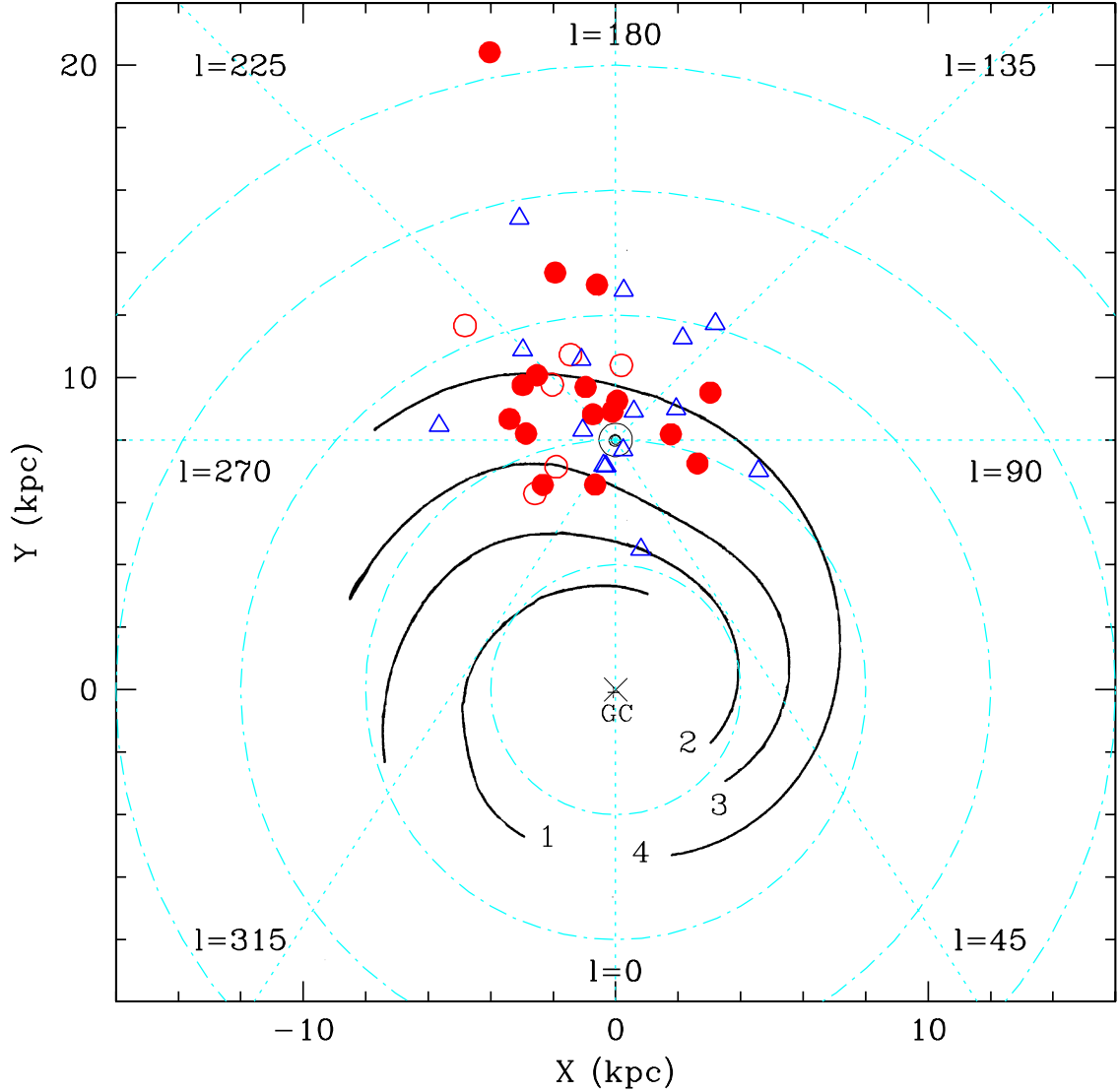}
\caption{Positions on the Galactic plane of the OCs in our sample; filled
circles  represent published OCs, open circles work in progress, and open
triangles the remaining clusters. We show the positions of the Sun and the
Galactic center, and a sketch of the spiral arms (respectively: 1 = Norma, 2 =
Scutum-Crux, 3 = Sagittarius-Carina,  4 = Perseus)}
\label{spiral}
\end{figure}

The clusters position in the Milky Way is shown in Fig.~\ref{spiral}, where
the four main spiral arms, as delineated by Georgelin \& Georgelin (1976) and
Taylor \& Cordes (1993) are also displayed. The Sun is assumed to be located
at 8 kpc from the Galactic center.  The clusters already analyzed and
published within the BOCCE photometric project are shown as filled dots,
whilst those with ongoing analysis (listed in the first lines of Table 2) are
represented by open circles; open triangles indicate the remaining clusters
(also listed in Table 2)  for which  the photometric observations have already
been performed.

\begin{figure}
\includegraphics[bb=40 175 475 700, clip=true, angle=0,scale=0.95]{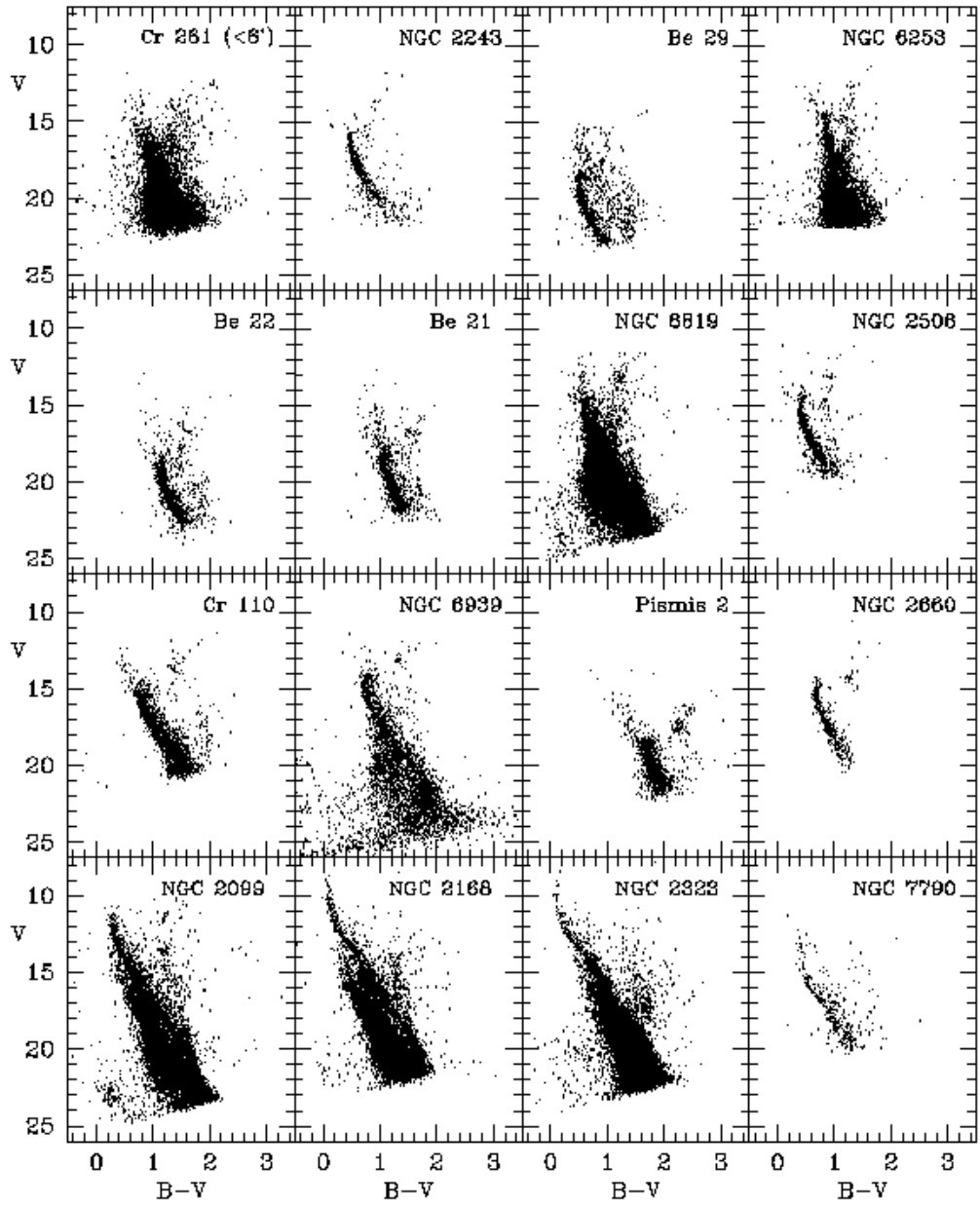}
\caption{V, B--V CMDs of the 16 OCs, shown in decreasing age order.}
\label{cmdbv}
\end{figure}

\begin{figure}
\includegraphics[bb=50 410 575 700, clip=true, angle=0,scale=0.95]{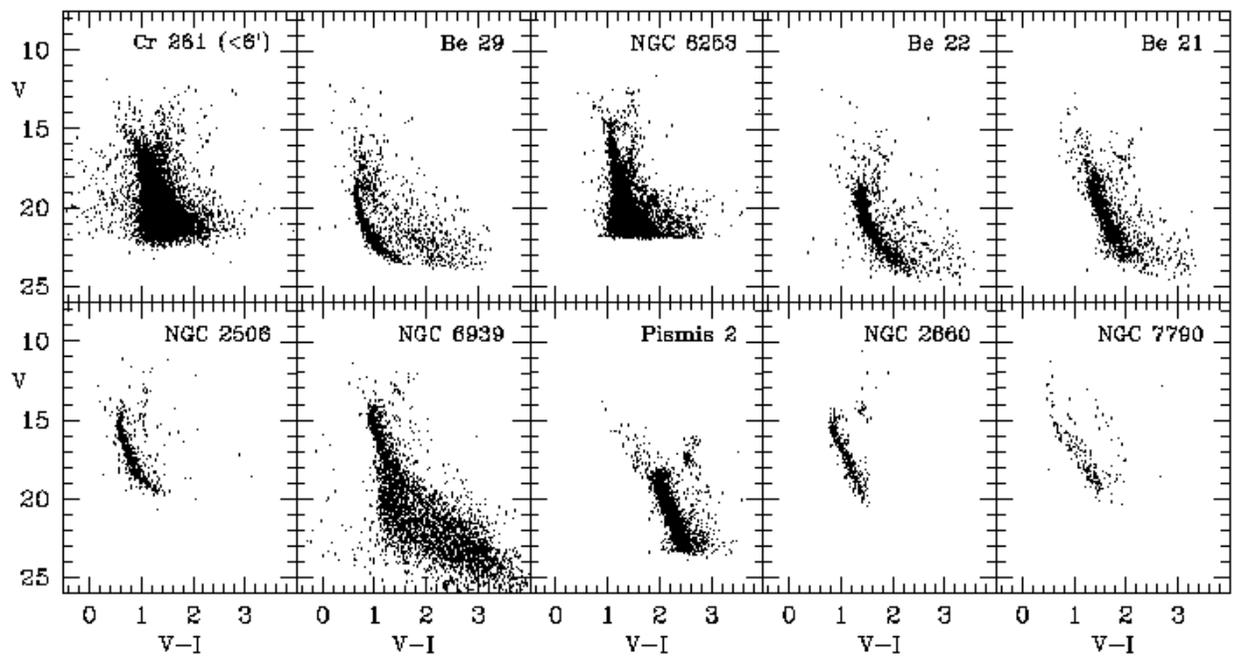}
\caption{V, V--I CMDs of the 10 OCs with I data, shown in decreasing age order.}
\label{cmdvi}
\end{figure}

Fig.~\ref{cmdbv} shows the empirical CMDs in the B--V color of the 16 clusters
analyzed so far. They are plotted in order of decreasing age (see Table 1 and
Section 4) rightwards from the top left panel. The corresponding CMDs in  V--I
for those clusters for which observations were performed also with the I 
filter are shown in Fig.~\ref{cmdvi}.

The appearance of the CMDs clearly reflects both the Galactic location of the
cluster and the telescope performances. Most clusters were observed with CCDs
mounted at 1 meter-class telescopes (the 1.5m Danish and the 0.9 Dutch
telescopes in La Silla, Chile,  the 1.5 m Bologna Observatory one in Loiano,
Italy, the Palomar 60 inches, USA).  The exceptions  can easily be recognized
from the depth of the CMDs: NGC  6819, NGC  2099, NGC  2168 and NGC  2323 come
from the CFHT open cluster survey (Kalirai et al. 2001a, 2001b, 2001c, Kalirai
et al. 2003) and  NGC  6939 from the Italian National telescope (TNG) on the
Canary Islands, both  being 4 meter-class telescopes. Finally, Be 22 and Be
29, while observed  with a similar diameter telescope (the ESO New Technology
Telescope on La Silla), are the farthest ones. 

Field stars contamination appears in all CMDs. Its prominence  depends of
course on the sky area sampled (from the 3.8 $\times$ 3.8 arcmin$^2$ of the
Dutch telescope, to the 42 $\times$ 28  arcmin$^2$ of the CFHT), but generally
speaking, the clusters seen in the direction of the Galactic Anticenter (e.g.
Be 21, Be 29 and Cr 110) are less affected by fore/background  contamination
than those located towards the center (NGC  6253 and Cr 261).  Furthermore,
the contamination is lower in the third quadrant ($180\degr \leq$ l $\leq
270\degr$) than in the others, with nearby systems like NGC  2099,  NGC  2168,
NGC  2323, NGC  6819 and NGC  6939, with galactic longitude between $\sim
74\degr$ and $\sim 221\degr$, as heavily contaminated as NGC  6253 and Cr 261.
This depends partly on the proximity of populous spiral arms and partly on the
presence of overdensities such as the disk warp and/or accreted satellites
streams (see Discussion).

\section{The application of the synthetic CMD method }

Age, reddening and distances of all clusters have been derived applying  the
synthetic CMD method (Tosi et al. 1991) to the empirical CMDs.  The best
values of the parameters are found by selecting those providing synthetic CMDs
with morphology, number of stars  in the various evolutionary phases and
luminosity functions (LFs) in better  agreement with the empirical ones. 

The synthetic CMDs are constructed via MonteCarlo extractions of mass - age
pairs, according to an assumed IMF, SF law, and time interval of the SF
activity. Each extracted synthetic star is placed in the CMD by suitable
interpolations on the adopted stellar evolution tracks and adopting the
Bessel, Castelli \& Pletz  (1998) tables for photometric  conversion to the
Johnson-Cousins photometric system. The absolute magnitude is converted  to a
{\it provisional} apparent magnitude by applying (arbitrary) reddening and 
distance modulus. The synthetic stars extracted for any {\it provisional}
magnitude and photometric  band are assigned the photometric error derived for
the actual stars of the  same apparent magnitude. Then, they are randomly
retained or rejected on the  basis of the incompleteness factors of the actual
data, derived from extensive artificial star tests. 

Once the number of objects populating the whole synthetic CMD (or portions of
it) equals that of the observed one, the procedure is stopped, yielding the
quantitative level of the SF rate consistent with the observational data for
the prescribed IMF and SF law.  To evaluate the goodness of the model
predictions, we compare them with: the observational LFs, the overall
morphology of the CMD, the stellar magnitude and color distributions, the
number of objects in particular phases (e.g., at the main sequence turnoff, MS
TO, on the red giant branch,  RGB, in the clump,  etc.). A model can be
considered satisfactory only if it reproduces all  the features of the
empirical CMDs and LFs. Given the uncertainties  affecting both the photometry
and the theoretical parameters (stellar evolution tracks included), the method
cannot provide strictly unique results; however, it  allows us to
significantly reduce the range of acceptable parameters.

In this way we derive the age, reddening, and distance of the cluster, and
indicate the metallicity of the stellar evolution models in better agreement
with the data.  In principle, the latter should be indicative  of the cluster
metallicity, but this depends significantly on some of the  stellar model
assumptions, such as opacities. 

\begin{figure}
\includegraphics[bb=170 150 550 695, clip=true, angle=0,scale=1]{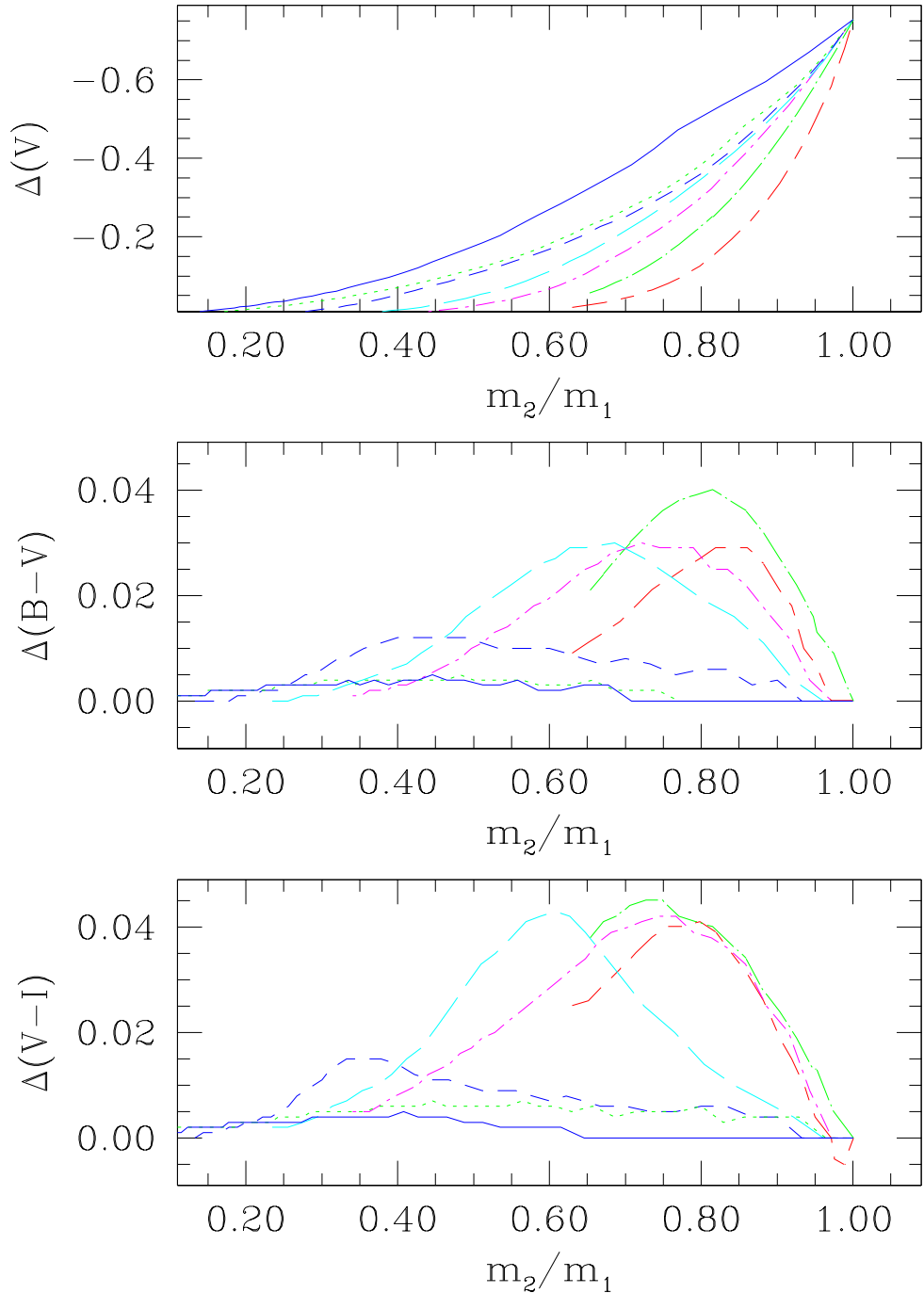}
\caption{Effect of unresolved binaries on the apparent magnitude and color as
a function of the mass ratio between the secondary and primary component of
the system and for various masses of the primary star.  Line symbols refer to
primaries of the following initial mass: 26 (solid), 14 (dotted), 4.5
(short-dashed), 2.55 (long-dashed), 1.76 (short-dash-dotted), 1.14
(long-dash-dotted), 0.95 \msun (short-dash-long-dashed). The shown variations
refer to solar metallicity MS stars from the Padova isochrones (Bertelli et
al. 1994).}
\label{binary}
\end{figure}

To estimate if, and how many, unresolved binary systems could be present in
the cluster, the synthetic CMDs are computed either assuming that all the 
cluster stars are single objects or that an (arbitrary) fraction of them are 
members of binary systems with random mass ratio. The effect of unresolved
binarism on the V magnitude and B--V color was first addressed by Maeder
(1974)  and more recently discussed by Hurley \& Tout (1998). Here we have 
re-calculated the effect also on the V--I color using the Padova isochrones
(Bertelli et al. 1994) and included it in the construction of the synthetic
CMDs. Fig.~\ref{binary} shows the adopted magnitude and color variations of MS
objects of solar metallicity as a function of the mass ratio  between the
secondary and the primary component of the  unresolved system.  Again, the
comparison of the resulting  morphology and  thickness of the synthetic MSs of
each cluster with the observed ones  allows us to derive information on its
most likely fraction of binaries.

The advantage of the synthetic CMD method over the classical isochrone fitting
to derive the clusters parameters is manifest: with the synthetic CMDs we take
into account both the number of stars predicted by the adopted stellar  tracks
in each evolutionary phase (directly related to the stellar lifetimes  in that
phase) and the effects of the photometry (errors, incompleteness  and image
blend) on the morphology and spread of the various CMD sequences. This clearly
provides a larger number of constraints and leads to a better selection of the
solution, if not to its uniqueness. For instance, an isochrone may appear to
well fit the data, but then the corresponding  distribution of the number of
stars on the MS or at its TO is completely at  odds with the empirical ones
and the case must be rejected, or be given a lower ranking with respect to
other models  (see e.g. Kalirai \& Tosi 2004). Moreover the blue and red edges
of the MS distribution are important tools to quantify both the reddening (its
mean value in the cluster direction and the  existence of differential
variations) and the presence of binary stars.

To test the effect of different input physics on the derived   parameters, we
have run the simulations of each cluster CMD with several different types of
stellar evolutionary tracks.  The adopted sets were  chosen because they
assume different prescriptions for the treatment of  convection and range from
no overshooting to rather high overshooting from  convective regions.    They
are thus suitable to evaluate the intrinsic uncertainties still  related to
stellar evolution models. The three types of stellar models applied to all
clusters are: the tracks of Ventura et al. (1998 and private communication,
hereafter FST)   with high ($\eta$ = 0.03), moderate ($\eta$ = 0.02) and no
($\eta$ = 0.0) overshooting;  the tracks of the Padova group (Bressan et al
1993, Fagotto et al. 1994, hereafter BBC) with overshooting; and the tracks of
the Frascati group (Dominguez et al. 1999, hereafter FRA)  with no
overshooting. The parameters of some of the clusters were also inferred,
in the original papers,
adopting the models computed by the Geneva group (first Maeder \& Meynet 1989
and then Schaller et al. 1992 and Charbonnel et al. 1993) and by Vandenbergh
(1985).

Some of the clusters of our sample were analyzed and published before the 
advent of the stellar models currently available and their parameters were 
then derived with the same method but with different stellar tracks than
clusters more recently studied. For such systems we have re-derived age,
distance and reddening with the version of the BBC, FRA and FST  models quoted
above, so that all the  results presented here have been obtained
consistently. For sake of homogeneity, also the fore coming analyses of the
remaining clusters of the BOCCE project will be performed with the same sets
of evolution tracks.

For those cases where the metallicities are not well constrained, we allow 
different values in our comparisons. We consider as solar metallicity tracks
those with Z=0.02, because they are the ones  calibrated by their authors on
the Sun, independently of the circumstance  that nowadays the actual solar
metallicity is suggested to be lower (see  Asplund et al. 2004). 

All models assume that the star formation activity has lasted 5 Myr (i.e., 
approximately an instantaneous burst, relative to the age of most of our
studied clusters) and that the stars were formed following a single slope ($x$
= 1.35) Salpeter's IMF over the whole mass range covered by the adopted tracks
(0.6 -- 100 $M_{\odot}$): simplistic assumptions which however don't affect
our results, as demonstrated by several checks made assuming both different
burst durations and IMFs.

\section{Ages, reddenings and distances from the CMDs}

Table 3 lists age (in Gyr), intrinsic distance modulus and reddening  derived
for each cluster with the synthetic CMD method using the  three different sets
of stellar models  described above.  Each triplet of parameters refers to the
best cases obtained with the BBC, FRA and FST models, for all the tested
metallicities. The differences in the derived quantities provide a
quantitative estimate of the intrinsic error attributable to the stellar
evolution tracks. 

Age is the parameter more affected by the theoretical uncertainties. Table 3
shows that there can be a factor up to 1.8 between ages derived with stellar
models with or without overshooting from convective regions. It is well known 
that, for a given mass, stellar models taking overshooting into account  have
larger cores and are therefore brighter than models without it. This makes
them appear as stars of slightly larger mass. Correspondingly, an object of
given luminosity, in the framework of overshooting models has  a lower mass
(and therefore an older age) than with  no overshooting models. Indeed, this
is what we find for the majority of our studied clusters: the BBC and FST
models provide ages older than the FRA models. The few exceptions to this rule
are due to uncertainties on the best fitting model metallicity. If we do not
consider the two oldest clusters Cr 261 and NGC 2243, for which the BBC models
appear to have problems (see next Section), the ages derived with the two
types of overshooting models, BBC and FST, are in fairly good agreement with
each other, since they  differ at most by 20\%.

Distance modulus and reddening appear to be much less affected by differences
in the adopted stellar models, specially if we take into account that the
available BBC, FRA and FST evolutionary tracks do not have the same nominal
metallicities. In the range of interest for open clusters, the BBC models are
available for metallicity Z = 0.004, 0.008, 0.02 and 0.05, the FRA models for
Z = 0.001, 0.006, 0.01 and 0.02, and the FST models for Z = 0.006, 0.01, 0.02
and 0.03. Hence, only the CMDs of  solar metallicity clusters could  be
compared with synthetic CMDs with the same metallicity for all types of
tracks. This leads to an uncertainty in the derived reddening, which
inevitably implies an uncertainty also in the distance modulus.  On the other
hand,  all  theoretical CMDs were transformed to the observational plane with
the same photometric conversion tables (Bessel et al. 1998), instead of 
directly taking values given by each group, thus  avoiding a further source of
uncertainty.

Since the BBC models have turned out to better reproduce the overall features
of the empirical CMDs of most of the clusters, we will adopt their resulting
parameters as reference values. 
of age, intrinsic distance modulus \mmm and reddening  \ebv attributed to each
cluster by the BBC stellar models from the application  of the synthetic CMD
method. The corresponding TO mass, model metallicity, distance d from the Sun
and R from the Galactic center are also given. In a few cases (see next
Section) the BBC evolution tracks were found to reproduce the clusters
observed properties less  satisfactorily than the FST or the FRA models. We
have flagged with a colon the values in Table 1 corresponding to these cases,
which are to be considered more uncertain. In the case of Cr 261, where the
BBC models are likely to have failed, Table 1 shows the values inferred with
the FST models. These values happen to be in agreement with those derived for
Cr 261 by Carraro, Girardi \& Chiosi (1999) via isochrone fitting with the new
Padova stellar models (Girardi 2000).

Detailed comparison with other analyses of  each cluster has been performed in
the individual papers, and we do not repeat it here; we only mention  the
recent work on NGC 2243 by Anthony-Twarog, Atwell \& Twarog (2005), since our
original paper is old, and this study constitutes an important improvement on
this cluster. As part of a larger project involving a series of key open and
globular clusters, they used Str\"omgren photometry to derive the cluster
parameters, arguing that intermediate-band photometry is able to reach more
precise results  than broad-band photometry, in particular for the
metallicity. NGC 2243 was targeted because it is old, metal-poor, and just
outside their proposed discontinuity near R$_{\rm GC}$ = 10 kpc.  They compare
their results with Bonifazi et al.'s  (1990), finding good agreement. This is
still true with our present analysis, since they find as best values \ebv =
0.055, [Fe/H] = $-$0.57, age = 3.8 Gyr, and apparent distance modulus 13.15
(see our updated parameters in Tables 1 and 3). They also rightly caution
that, since the resulting estimates are tied  to the adopted set of
evolutionary tracks,  it is safer to work on  a relative scale to derive the
properties of a sample of clusters. They also compare Be 29 to NGC 2243, again
finding very good agreement with our determinations.

\section{The cluster CMDs as stellar evolution tests}

When the BOCCE project started many years ago (Romeo et al. 1989), the 
stellar evolution models available in the literature were still significantly 
affected by uncertainties in the treatment of the convective regions, in the
adopted opacity tables, in the normalization to observed systems. The
photometric conversions from the theoretical effective temperature --
luminosity plane to the empirical color -- magnitude plane were very uncertain
too and often created ad hoc for each grid of stellar tracks, to let it
reproduce the colors of observed objects.  Moreover, most stellar models were
computed for relatively small mass ranges, and it was therefore quite
difficult to set up a homogeneous set of models covering the whole mass range
of interest for open clusters (i.e. low and intermediate mass stars from 0.6
to 8--10 \msun).  As a consequence, several of the adopted models were unable
to properly reproduce all the CMD features of our observed clusters; first of
all, the color of the red giant branch (RGB) and the morphology of the MS TO.

The three types of stellar models (BBC, FRA and FST) adopted now are those
that have proven most successful so far in reproducing the observational
properties of the clusters. Nonetheless, none of them reproduces perfectly all
the CMD features. The differences between predicted and observed  features
sometimes may be attributed to uncertainties in the data\footnote{ for
instance, Twarog, Anthony-Twarog \& De Lee 2003 have suggested the existence
of a color term in the calibration of our photometry of  NGC  6253, comparing
it to other analyses of the same cluster; for this cluster there also appear
to be quite large zero point shifts among available photometries.  Anyway, the
values they derive for this cluster are in reasonable agreement with our
findings: metallicity definitely higher than solar, age between 2.5 and 3.5
Gyr, reddening 0.26 and apparent distance modulus between 11.6 and 12.2.}, but
in most cases they are the systematic consequence of the stellar model
assumptions.

The FST models are very instructive in this respect because, for any given
mass and metallicity, they provide three sets of tracks computed with the
same assumptions except the amount of overshooting. By comparing the synthetic
CMDs based on these three different amounts of overshooting ($\eta$=0.0, 0.02
and 0.03) with the empirical CMDs of our clusters we have found (Ventura et
al. in preparation) that the models with overshooting are in much better
agreement with the data than the models without it. Even for Cr 261, where
core overshooting is in practice not effective because at that old age (6 Gyr
with the FST models)  MS stars have low mass and hence radiative cores, the
$\eta$=0.02 models are slightly better than the $\eta$=0.0 ones. The only
exception to this rule is NGC  2243 (which is also rather old,  with an age of
4 Gyr with the BBC models and 2.9 with the FST ones),  where all the FST
models fail to provide a good fit to the empirical CMD, but the ones with
$\eta$=0.0 are less inconsistent than the others. In all the other cases with
TO mass large enough to let the stellar core be convective (and overshooting
be effective), the $\eta$=0.0 models  either do not populate correctly the
clump and/or the TO region, or do not  show the correct MS curvature. 

It is interesting that, while the need of overshooting in the FST models is
evident, it is almost impossible to systematically choose between  $\eta =
0.02$ or 0.03 from our data. For some clusters we don't find  significant
differences in the agreement between empirical and synthetic CMDs switching
from 0.02 to 0.03; for other clusters (Be 29, Cr 110, NGC  6939, Pismis 2) we
find that the $\eta = 0.02$ models better reproduce the data; but for yet
other clusters (NGC  6819, NGC  2660, NGC  2168 and, possibly, NGC  2323) we
find  instead a better fit with the $\eta = 0.03$ models. Looking at the
values listed in Table 1, one might be tempted to find hints for an
anti-correlation  between the age of the cluster and the amount of
overshooting, but we think that the sample is still too small, the
uncertainties too large and the relation too loose to support this suggestion.

One intriguing aspect of stellar evolution models concerns the slope of the 
lower MS. Sometimes the photometry is not accurate enough to allow for a safe
definition of the MS shape and curvature at faint magnitudes,  but there are
cases in which the photometry is reliable and the synthetic CMDs (and
isochrones) do  not reproduce them perfectly. This problem has been recently
addressed by Grocholski \& Sarajedini (2003), who presented a comparative
study of isochrone fits to CMDs in the visual and infrared bands. They took
five sets of theoretical models and compared them in the luminosity -
temperature plane.  They fit the observed CMDs of six open clusters, of
various ages and metallicities, with the five sets of isochrones and found a
variety of  inconsistencies on the lower MS slope. Their results are somewhat
affected   by the fact that  they use the isochrones as transformed from the
theoretical plane  (luminosity and temperature) to the observational one
(magnitudes and  colors) by each group of stellar modelers, i.e. following 
different  prescriptions and introducing differential variations on the slope
itself. In our case, this differential (and unknown) effect is absent, because
we transform all the theoretical models with the same conversion tables.
Still,  we find the same kind of problems as described by Grocholski \&
Sarajedini: when the observed CMDs are deep enough, the fit to the lower main
sequence is often (but intriguingly not always) bad. This happens in general
around M$_{\rm V} \simeq $ 8  for the visual colors, and is attributed by 
Grocholski \& Sarajedini to some ingredient still missing in the model
atmospheres for low mass stars. Indeed, this is likely to be at least one of
the causes. For instance, the circumstance that most photometric conversions
are applied under gray atmosphere assumptions, which are likely not to be
applicable to stars with M$_{\rm V} \geq $ 8 is a probable source of
inconsistency. Whatever the reason, there is no single set of tracks able to
match all CMDs in all  colors, at all ages, and for all metallicities.  All of
them behave quite well in some cases, but not in others.

With our sample of OCs, we have found that the FST models in some cases 
predict too steep MS slopes at the lower masses, in spite of being usually 
able to reproduce the observed features of the upper MS and TO better  than
other models. In fact, we have found that, while reproducing the  colors and
magnitudes of the upper MS, the faint end of the FST synthetic MS is too blue
(and bluer than that of the BBC and FRA models) for NGC  2243, NGC 6819 and
NGC  6939. In a few cases (e.g. NGC  2660) the FST models overpredict the
number of faint  MS objects more than other stellar tracks (i.e. more than
attributable to evaporation effects).

The FRA models are the only ones able to well reproduce the MS curvature of
NGC 6819 and Pismis  2, and fit perfectly the older cluster Cr 261.  On the
other hand, they often overpopulate the post-MS phases and present a  too
hooked TO morphology at young and intermediate ages (see e.g. NGC 2099,  NGC
2168, NGC 2323 and NGC 6819 in Kalirai \& Tosi 2004), which make them less 
viable than other tracks for clusters younger than $\sim$2 Gyr.

The BBC tracks turn out to be those in overall better agreement with the
observed CMDs. In other words, they steadily provide a good fit to the data of
(almost) any age and metallicity, even if they may appear to reproduce 
detailed features, such as MS gaps and bumps, TO shape, clump numbers,  in a
less strikingly good manner than other models.  The only case where the BBC
tracks have not led to a convincing result is Cr 261. For that cluster we
(Gozzoli et al. 1996) found an age of 11 Gyr with the BBC models, much older
than with tracks without overshooting, as if overshooting were strongly active
in the cluster stars, despite their low mass (the TO mass is about 1.1 \msun.)
which implies a radiative and not a convective core. We have reexamined  its
case, with the FRA and FST models, not available at the time of that study, 
and found a very good fit to the observational CMD for an age of 5 Gyr with
the FRA models or 6 Gyr with the FST models (independently of the adopted
amount of overshooting, as expected, since overshooting should not be active
in low mass stars). The problem with the BBC tracks (and isochrones, Bertelli
et al. 1994) is that at ages of 6 to 9 Gyr they show a hooked TO, which is
usually appropriate for younger systems and does not reproduce the smooth
tighter morphology of the TO of Cr 261. At older ages, the BBC TOs recover the
proper shape, but then the clump luminosity becomes excessively bright for Cr
261.  It is interesting to note that the new Padova isochrones (Girardi
2000) present  TO shapes much more similar to those of the FST
and FRA  models (and of Cr 261, see also Carraro et al. 1999 and their 
fig.1) at these ages. We thus feel that the BBC stellar models might have had some problems
in the TO evolutionary phases of low mass stars, subsequently overcome in more
recent computations.

\section{Field contamination, reddening and binary systems}

A safe derivation of cluster age, reddening and distance from CMDs requires
not only reliable photometry, with  accurate calibration, but also a careful
analysis of the back/foreground contamination and as much as possible
information on the actual cluster membership of the measured stars. In fact,
the shape and the luminosity of the corner stones for the parameters
derivation (upper MS,  TO  and clump) may be significantly altered  by the
presence of line of sight intruders. Unfortunately, studies of radial
velocities or proper motions of the individual stars are available only for a
few open clusters and often the only means to estimate cluster membership is
by comparison with nearby blank fields. De-contamination is even more crucial
if one aims at deriving also the cluster initial mass function or the star
formation rate. We thus emphasize the importance of acquiring the photometry
of appropriate blank fields, even if this requires additional observing time
and a careful choice of the field position.  The case of the beautiful CMDs
from the CFHT open cluster survey (Kalirai et al. 2001a, 2001b, 2001c, Kalirai
et al. 2003) is illustrative. Despite the large field of view of the CFH12K
mosaic camera  (42$^{\prime} \times 28^{\prime}$) Kalirai \& Tosi (2004) have
found that in the case of NGC ~2168 and NGC ~2323 many cluster members are
still present  even at the edges of the observed field:  the lack of
observations of an external field prevents  any reliable estimate of the 
back/foreground contamination.  

Open clusters are affected by dynamical friction with the Galactic disk and by
stellar evaporation, so that stars initially  members can be found at rather
large distances from the cluster center. An example is given e.g. by Be 22 (Di
Fabrizio et al. 2005, see figs. 4 and 5), for which we acquired a comparison
field at a distance of 30 arcmin from the cluster center that still shows a 
main sequence similar to that of the cluster. Since this MS is  not present
in  the Galactic synthetic models of the field disk population by Robin et
al.  (2003), its most likely interpretation is that cluster members are
present even at such large distance. This makes it additionally uncertain to
infer membership by statistically  subtracting the objects found in adjacent
fields. For this reason,   we have generally restricted our derivation of the
cluster parameters to the  CMDs of their most central regions, where field
contamination is smaller. The synthetic CMD method, by providing the
theoretical number of stars in all evolutionary phases, allows to easily
individuate the clusters where  evaporation is more likely to have occurred.
Clusters whose fainter main sequences are systematically underpopulated with
respect to those of the corresponding  synthetic CMDs are robust candidates
for evaporation, once the reliability of the incompleteness factors is
guaranteed by extensive and appropriate artificial star tests. In our sample,
we have found that NGC 6253, NGC 6819, Be 29 and Cr 261 have faint MSs 
systematically less populated than the corresponding ones predicted by all the
stellar models. We thus consider these clusters as having suffered significant
evaporation of low mass stars during their life. These  four systems are among
the oldest ones of the provisional BOCCE sample and have had more time than
the others to loose an appreciable fraction of their stars. It is interesting
to notice, though, that Be21 and NGC 2243, which are also older than 2 Gyr, do
not show evidence of evaporation. NGC 2243 is by far the OC more distant from
the Galactic plane in our entire sample, and this may explain it; no simple
explanation is available for Be 21.

\begin{figure}
\includegraphics[scale=0.9]{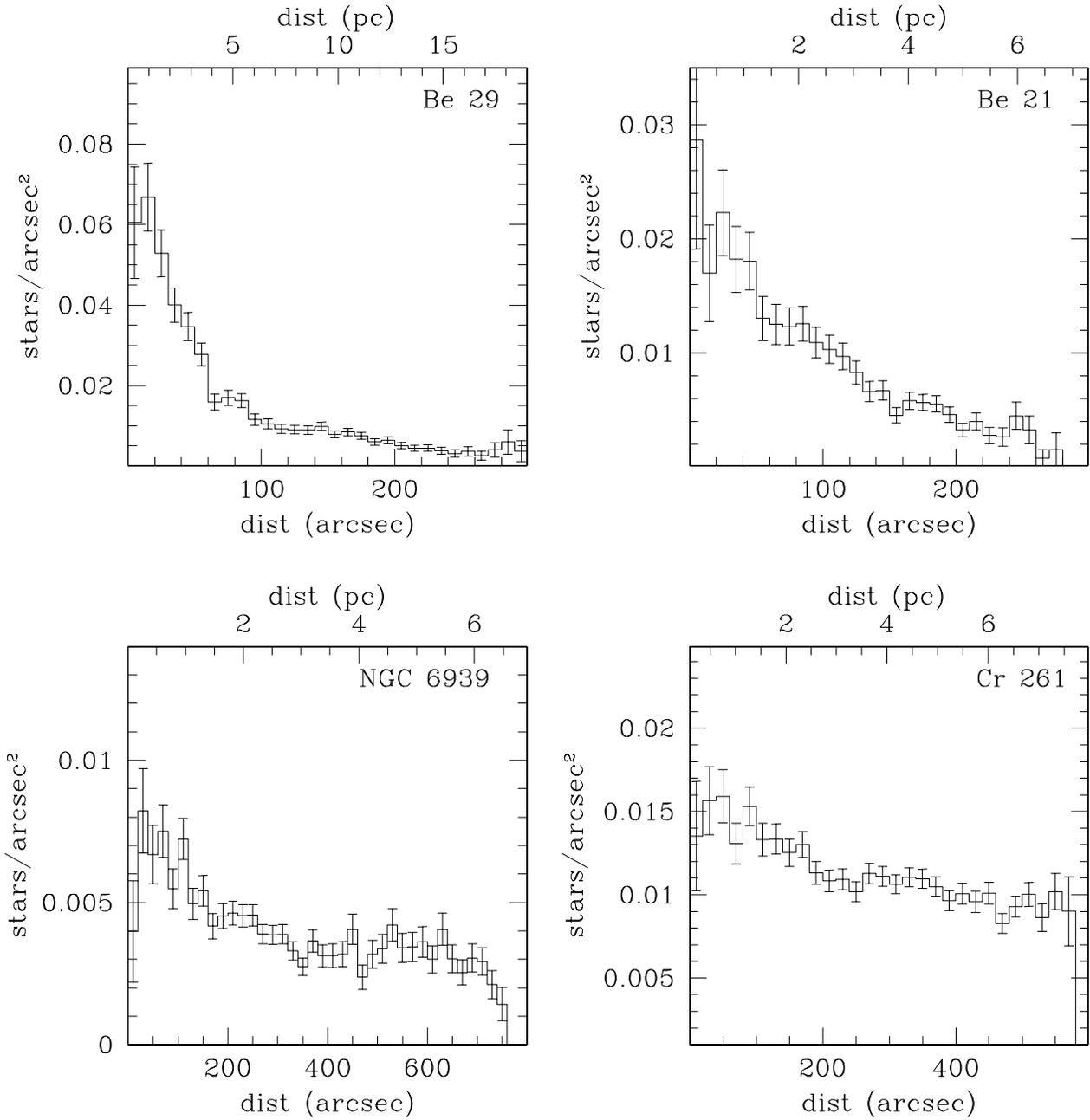}
\caption{Projected density distribution for Be 29, Be 21, (both computed with 
a bin 10 arcsec wide), NGC 6939, and Cr 261 (both with bin = 20 arcsec).}
\label{area}
\end{figure}

We have not generally tried to determine the extension of our OCs, partly
because it was outside the main goal of our project, partly because our field
of view is often so small as to make it impossible.  Only for the farthest
clusters (Be 29, Be 22 - already discussed in Di Fabrizio et al. 2005 - Be 21)
or for the few observed with a larger field of view (NGC 6939, Cr 261) the
exercise if worthwhile. The radial extent of the 4 OCs observed with the CFHT 
has already been determined (Kalirai et al. 2001b,c; 2003): NGC 6819 has a
radius of 9.5 arcmin, NGC 2099 of 13.9 arcmin, while the determinations for
NGC 2323  (about 15 arcmin) and NGC 2168 (more than 20 arcmin) are more
uncertain, since the clusters seem to fill the entire field of view. We
computed  the radial star density distribution for Be 29, Be 21, NGC 6939, and
Cr 261.  We show in Fig. \ref{area} the projected density distribution  (in
annuli 10 or 20 arcsec wide) for the 4 OCs, giving distance form the centers
both in arcsec and in parsec. The only one for which a flattening of the
distribution is indisputable, so we can determine where the cluster
disappears, is Be 29 (with a radius less than about 3 arcmin, in good
agreement with the   value in the  Dias et al. 2002 catalogue). The
distribution for Be 21 is still decreasing at our field's edge, and the same
seems to be the case for NGC 6939, where we chose only stars brighter than V =
22, to avoid problems with an excessive field stars fraction and
incompleteness of our photometry.  Finally, Cr 261 (where we considered only
stars brighter than V = 20) seems to constitute a small fraction of the
observed field, as already found by Gozzoli et al. (1996).

As many disk objects, our clusters are generally affected by significant 
reddening.  However, only few of them clearly suffer also from differential
reddening. The synthetic CMD method, by distributing the stars in the CMD
according not only to the evolutionary phase, but also to photometric errors
and fractions of presumed binary systems, allows one to quantify differential
reddening  more safely than with simple isochrone fitting. A  varying
reddening widens the whole MS by a given amount $\Delta$\ebv, while unresolved
binaries produce a secondary sequence redder/brighter than  the single stars
MS, intersecting it above the TO region. In practice, differential reddening
and binarism are distinguishable from each other because the former implies a
smooth spread of the MS from the TO down to its faintest portions, whilst
unresolved binaries, even if distributed with random mass ratios, imply a
split of the MS and the appearance of a bright appendix (up to 0.75 mag
brighter, corresponding to equal mass binaries) on  top of the TO.  Based on
these arguments, we have found that  Be 21 and Pismis 2 are the only systems
of our current sample clearly affected by differential reddening (0.04 $\lsim
\Delta$\ebv $\lsim$ 0.1). Work in progress (Bragaglia et al. 2005, in
preparation) on the photometric data and the spectra of three stars in NGC
3960 indicate also for this cluster some differential reddening, confirming
the finding by Prisinzano et al. (2004). 

All the clusters, instead,  seem to contain unresolved binary systems, with
estimated percentages varying  between 15\% and 60\%.

NGC 2099, NGC 2168, NGC 2323 and NGC 6819, in addition to the binary sequence,
show MS widths larger than attributable to the measured photometric errors
(Kalirai \& Tosi 2004). This could be due to a differential reddening of the
order of 0.01 mag and/or to internal metallicity spread. However, in these
cases we rather tend to ascribe the MS width to the difficulty in calibrating
the different CFH12K CCD chips (McCracken, private communication). 

\section{Discussion and preliminary results}

\begin{figure}
\includegraphics[bb=40 175 475 700, clip=true,angle=0,scale=0.95]{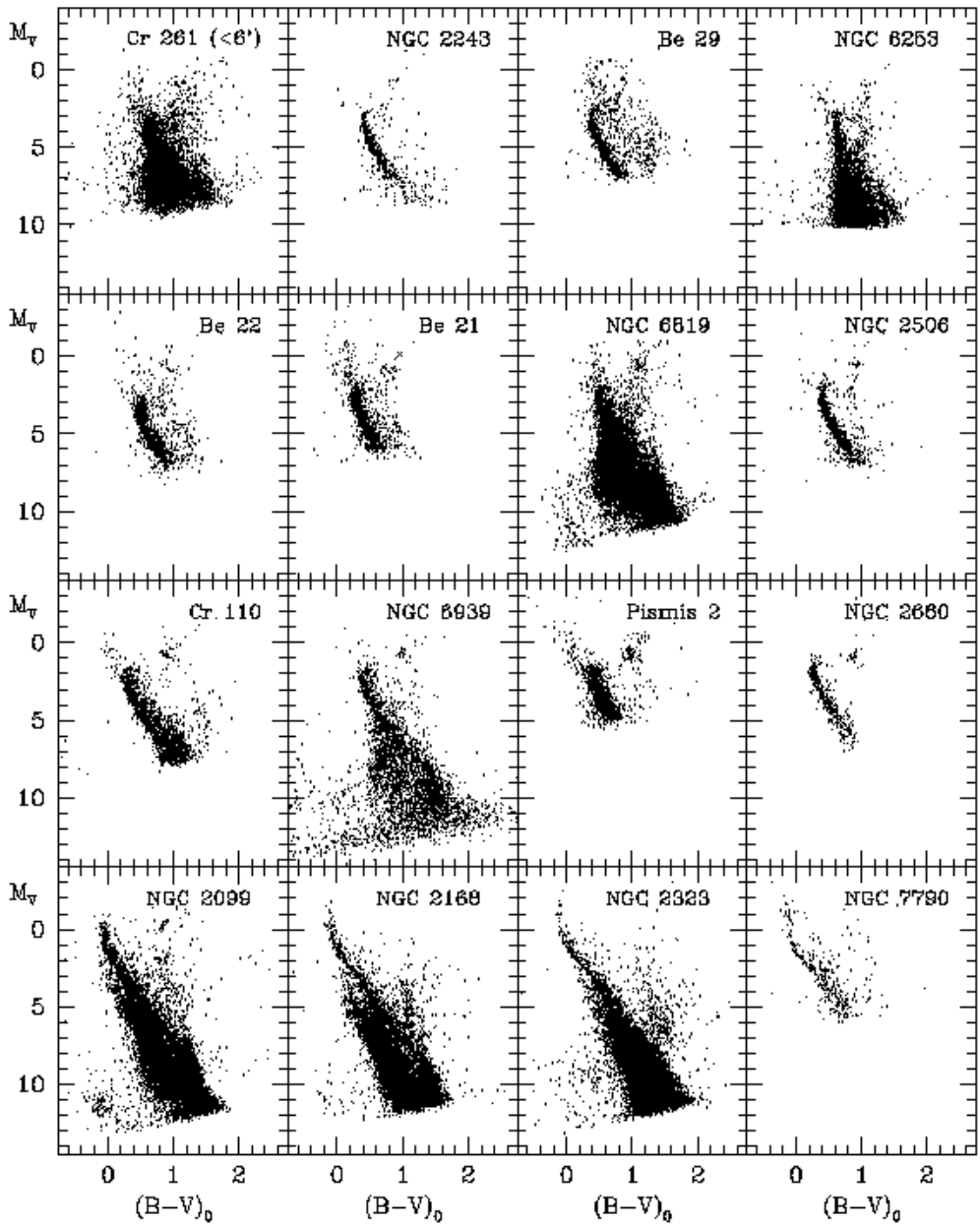}
\caption{CMDs for the 16 OCs in absolute magnitude and intrinsic color.}
\label{cmdbvo}
\end{figure}

With the derived reddening and distance modulus, the observational CMDs  can
be plotted as absolute magnitude vs intrinsic color (Fig.~\ref{cmdbvo}). In
this reference frame, the TO brightness obviously anticorrelates with the
cluster age, while the clump luminosity at first sight is roughly constant, 
although actually inversely proportional to the chemical abundance. At first 
order approximation, the magnitude difference, $\Delta$V, between TO and
clump  is thus an age indicator. A safe use of this indicator requires a
careful and homogeneous definition of the reference  TO and clump points,
which  are affected by photometric uncertainties,  including blending of
unresolved stars, small number statistics, and back/foreground contamination.
In practice, the identification of the reference points  differ from one author
to the other. For instance,  Twarog \&  Anthony-Twarog (1989) define the
morphological age  ratio (MAR), reddening independent and almost  metallicity
independent,   as the magnitude  difference between the brightest TO point
and the clump,  divided by the color  difference between the RGB color at the
clump luminosity and the bluest TO  point. Janes \& Phelps (1994) adopt as
morphological age index (MAI) the magnitude distance from the red clump of the
inflection point between the  TO and the base of the RGB, as did more
recently   Salaris, Weiss \& Percival (2004). Carraro \& Chiosi (1994) assume
that the reference TO luminosity is 0.25 mag fainter than observed in the CMD,
because of the possible presence of unresolved binaries which brighten the
apparent TO.

The oldest cluster of our provisional sample is Cr 261, whose age has been
conservatively estimated with the FST models, due to the problems with the BBC
tracks discussed in the previous Section. Its $\Delta$V is 2.6. The next older
systems are in the interval 4 -- 3 Gyr   (Be 29, NGC ~2243 and NGC ~6253),
with $2 \lsim \Delta$V$\lsim 2.5$. The  clusters younger than 1 Gyr (NGC
~7790, NGC ~2323, NGC ~2168, NGC ~2099 and NGC ~2660) all have $\Delta$V
between 0.2 and 0.5 and both the TO and the clump are much more difficult to
individuate, because of the small number of cluster members in these
evolutionary phases.

\begin{figure}
\includegraphics[bb=160 140 480 710,angle=0,scale=1]{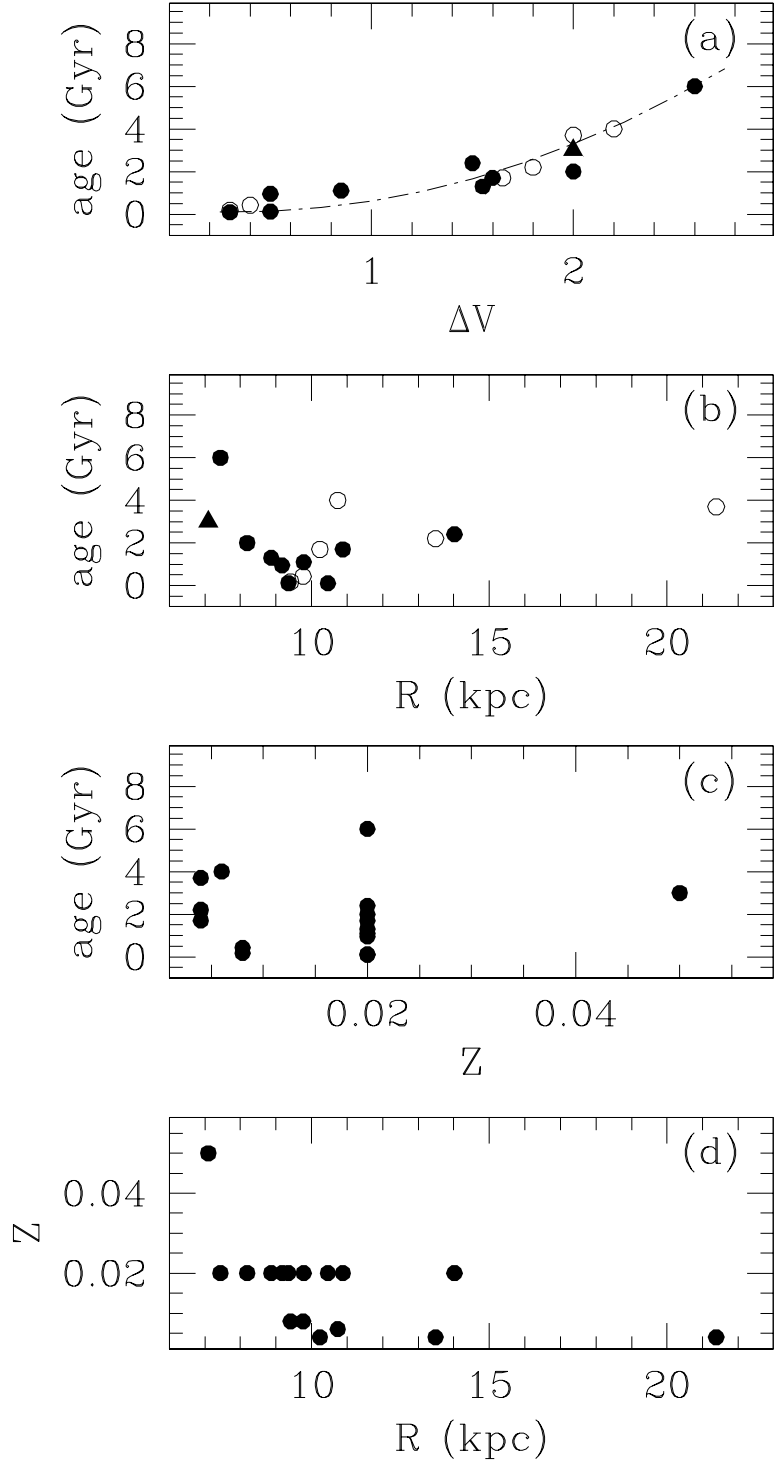}
\caption{For the clusters listed in Table 1 we show the distributions of:
(a) age vs magnitude difference between TO and clump (filled circles, open
circles, filled triangle indicate Z solar, subsolar or supersolar,
respectively), with a line representing our preliminary relation (see text);
(b) age vs Galactocentric distance (same symbols);
(c) age vs model metallicity;
(d) model metallicity vs Galactocentric distance.}
\label{rel}
\end{figure}

Fig.~\ref{rel}(a)  shows that $\Delta$V seems well correlated with the cluster
age derived with our  method, and the line represents a preliminary
calibration of this relation.
Although it appears as a very  reasonable approximation, we will not make use
of it in its present  form:  these 16 clusters are still too few to give a 
reliable calibration between $\Delta$V and age because  they were not chosen
for this purpose, and do not sample the entire  relevant age range (see e.g.
Salaris et al. 2004 for a discussion on their choice of calibrating clusters).
In particular, we miss the oldest OCs, and a definitive calibration will have
to wait until at least Be 32 (age $\sim$ 6 Gyr), Be 17 and NGC 6791 (both with
age $\sim$ 10 Gyr) will be properly analyzed (for the first two clusters work
is already in progress).  Furthermore, precise metallicities are, or will soon
be, available for (most of)  the clusters in our sample, either from our work
(e.g., Cr 261: Carretta et al. 2005; Be 17, Be 32, and NGC 6791 for which we
have acquired high resolution spectra) or from the literature (e.g., Be 29:
Carraro et al. 2004; Be 17: Friel, Jacobson, \& Pilachowsky 2005).  This is
important to test the metallicity dependence of the $\Delta$V - age relation.
As explained e.g., in Salaris et al. (2004), isochrones of different
metallicity have different values of $\Delta$V for the same age, since the
luminosities  of the TO and the RC depend in different ways on age and metal
abundance.  In particular, lower metallicities should show a larger $\Delta$V
for the same age.  In our case the situation is confused: we have three
clusters with the same $\Delta$V  but the age difference does not seem to
correlate univocally with the metallicity. Larger samples, and better defined
metal abundances, have to be employed to decide whether and how this influences
the calibration of this useful relation.

From the derived distance moduli and reddenings we calculated  the Galactic
position of the clusters, used to draw Fig.~\ref{spiral}. NGC ~6253 is the
system closer to the Galactic center and Be 29 the farthest one. To better
visualize the relations (or lack thereof) between age, Galactocentric distance
and metallicity, we have plotted these quantities in Fig.~\ref{rel}(b,c,d). We
recall, however, that the metallicity derived here is that of the BBC models
in better agreement with the cluster CMDs and should not be taken as a precise
measure of the actual cluster metallicity. To safely determine the cluster
chemical abundances accurate spectroscopic studies are necessary. Bearing this
caveat in mind, Fig.~\ref{rel}  shows that the only relation clearly existing
between the derived quantities is the metallicity dependence on Galactocentric
distance [Fig.~\ref{rel}(d)], while there is no apparent relation between age
and metallicity. If we don't take the outermost point into account, we cannot
exclude a dependence of the cluster age on Galactocentric distance; however,
the spread in the second panel from top is quite large and we need a much
larger sample to better pin point this relation (or lack thereof). Similar
results were already found and discussed by Janes \& Phelps (1994) and Phelps,
Janes \& Montgomery (1994), with a much larger  sample of clusters, but more
uncertain parameter derivation.  Carraro et al. (1998) argued that, after
correcting the cluster metallicities for the radial gradient effect,  their
clusters do show an age-metallicity relation similar to that inferred from
field stars in the solar neighborhood. However, their corrected 
age-metallicity distribution (their fig.8) looks as scattered as ours 
[Fig.~\ref{rel}(c)], with no apparent trend. Yong et al. (2005) find no
correlation between age and metallicity in their compiled sample of OCs. 

\begin{figure}
\includegraphics[bb=140 280 515 710, angle=0,scale=1]{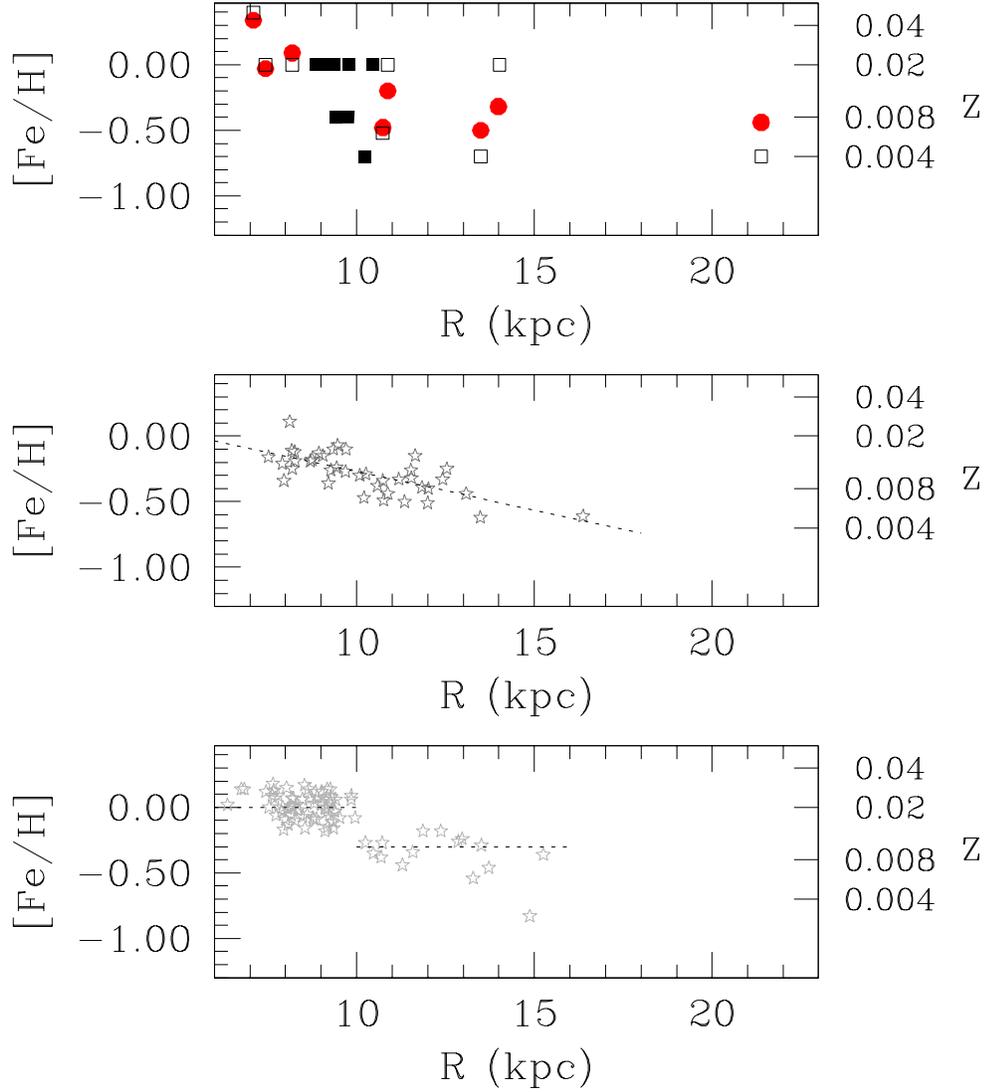}
\caption{ 
Galactocentric metallicity distribution. Here we show [Fe/H] instead of Z, but
on the right side of each figure we indicate the level of a few representative
Z values for immediate comparison. Top panel: our sample, where filled squares
are photometric metallicities, and filled dots are  [Fe/H] based on high
resolution spectroscopy while open squares are the corresponding photometric
values. Middle panel: Friel et al.'s (2002) data, with the gradient based on
their data, which has a slope of -0.059 dex kpc$^{-1}$. Bottom panel: Twarog
et al.'s (1997) data, that they interpret as a step distribution, with the two
metallicity levels indicated by the dashed lines:  [Fe/H] = 0 for R $<$ 10
kpc, and $-0.30$ for R $>$ 10 kpc.}
\label{grad}
\end{figure}

Do we really see a radial abundance gradient in the Galactic disk?  Fig. 
\ref{grad} shows our data\footnote{
We have translated our photometrically determined Z values to the corresponding
[Fe/H] values, for more immediate comparison with the two other samples. We
also plot here the available [Fe/H]'s based on high-resolution spectroscopy,
that are not (yet) on a  thoroughly homogeneous scale, since three of them
come from literature (see Table 4). 
} 
compared to the homogeneous samples by Friel et al.
(2002) and Twarog et al. (1997).  As already mentioned in the Introduction, 
the latter, on the basis of  their re-homogenization of OC metallicities,
presented the hypothesis that the true original metallicity distribution in
the Galactic disk is not  described by a simple linear gradient, but is  a
step distribution, with  clusters inside  R$_{\rm GC}$ = 10 kpc  with solar
metallicity  (and a dispersion of $\pm$ 0.1 dex) and those outside  with a
metallicity 0.3 dex lower, with a similar dispersion (see Fig. \ref{grad},
lower panel, which uses data from Twarog et al. 1997). They discussed at
length the different metallicity distributions of clusters and field stars and
suggested that the differences are mostly due to the much larger orbital
diffusions of individual stars. Corder and Twarog (2001)  modeled the
evolution both of a linear and a step distribution in metallicity for field
stars and open clusters, and found that a step distribution does appear as a
linear gradient if orbital diffusion is the dominant factor, like for field
stars. For open clusters, instead, evaporation  dominates, allowing to
preserve the original discontinuity. The large sample  in Twarog et al. (1997)
was dominated by youngish clusters (age $<$ 2 Gyr) and the original step
distribution (which still survives) could be recognized. Corder \& Twarog
(2001) attribute the linear gradient found e.g. by Friel (1995) to the too
small radial baseline, with most of clusters near the discontinuity (in the
range 8-12 kpc). They suggest to concentrate efforts on objects much nearer
to, and farther from, the Galactic center to avoid this effect.

Our clusters cannot prove or disprove this description yet: we still have too
few objects at any given  R$_{\rm GC}$, we are biased towards large ages, and
our homogeneous metallicities come from the evolutionary tracks. Our inner
clusters have predominantly solar metallicity, and of the three OCs definitely
outside the 10 kpc limit (remember that we are using  R$_{\rm GC,\odot}$ = 8
kpc), two have  [Fe/H] $\sim$ -0.4 (from spectroscopic analyses) and one has
Z=Z$_{\odot}$ (Be 22, from the synthetic CMDs method; see however Villanova et
al. 2005).  More clusters  at these Galactocentric distances should be studied
in detail and a more precise abundance determination has to be performed. We
list in Table 4 the presently available metallicities based on high resolution
studies; 5 were determined by our group and 3 come from literature, so they
are not yet on a homogeneous scale.

In any case, a simple single-slope gradient cannot be the best answer: our
present sample contains the very distant cluster Be 29, which  has abundances
higher than what is given by a linear extrapolation of the gradient found for
OCs nearer to the Galactic center (see Fig. ~\ref{grad}, upper and middle
panels). The same is true for Saurer A  with R$_{\rm GC} \simeq$ 19 kpc and
[Fe/H] = $-$0.38 (Frinchaboy \& Phelps 2002,  Carraro et al. 2004). The fact
that inner and outer disk clusters do not seem to follow the same radial
abundance distribution was already noticed by Carraro et al.  (2004), who
proposed two alternatives: a different chemical evolution history for the
outer disk (e.g. Chiappini et al. 2001) or an extragalactic origin of Be 29
and Saurer A\footnote{ They associated Be 29 to the Monoceros Ring, but not
Saurer A, at variance with Frinchaboy et al. (2004, 2005) }. Yong et al.
(2005), combining literature data with high-resolution spectral analysis of
M67 and five outer disk OC's (Be 20, Be 21, Be 29, Be 31, and  NGC  2141),
also argue that open clusters show a linear metallicity gradient in the
regions within 10 -- 12 kpc from the Galactic center, but have a roughly
constant metal content from there outwards. Their favored interpretation is
that the outer clusters are the result of star formation triggered by a major 
merger event. Direct capture of alien clusters from accreted dwarf galaxies
seems unlikely, since the relatively high [$\alpha$/Fe] ratios measured in
their  five clusters are not consistent with those measured in nearby dwarf
galaxies (Venn et al. 2004 and references therein).

Crane et al. (2003) and Frinchaboy et al. (2004) connected several open and
globular clusters to the Galactic Anticenter Stellar Structure (GASS; e.g.,
Newberg et al. 2002, also called Ring, or Monoceros Ring). Bellazzini et al.
(2004) proposed a few open clusters to be linked to the Canis Major dwarf
(CMa) that may (Martin et al. 2004), or may not (Momany et al. 2004) be a real
accreted satellite, and that may perhaps be (Martin et al. 2005), or not be
(Pe\~narrubia et al. 2005) the cause of the GASS. The two groups have
different approaches. Crane et al. (2003) and Frinchaboy et al. (2004, 2005)
use position and radial velocity of galactic clusters to see if they lie along
the GASS, while Bellazzini et al. (2004) look at the CMDs of clusters at about
the same position and distance of the CMa system, and compare the cluster and
surrounding field population with the CMa one. 

A few clusters in our sample have been considered  to be associated to CMa or
to the GASS, on the basis of their CMD and/or radial velocity. For  To 2
(Frinchaboy et al. 2004, 2005; Bellazzini et al. 2004) and  Be 29 (Frinchaboy
et al. 2004, 2005) there is a strong positive indication. Be 22, on the
contrary,  has been discarded both by our work  (Di Fabrizio et al. 2005) on
the basis of a comparison between the CMDs of the central and control fields
and that of the CMa overdensity, and by Frinchaboy et al. (2005) because its
radial velocity does not fit the GASS longitude-velocity trend (this remains
true also considering the slightly lower radial velocity determined by
Villanova et al. 2005). In the case of Be 29 we have tried to apply both
methods to our data (Tosi et al. 2004): the galactic coordinates and radial
velocity (measured by Bragaglia, Held \& Tosi 2005, and by Carraro et al.
2004) put it right in the GASS  (see also fig. 1 in Frinchaboy et al. 2005),
but our field of view is too small to give a definite answer about 
similarities between Be 29 and CMa or dissimilarities with the expected
galactic disk population based on Robin's et al. (2003) model. 

All the other clusters of our present sample are too close to the Galactic
center and/or have radial velocities that exclude the association with the
Ring (see Table 4). We have anyway checked that no unaccounted component was
present in our CMDs by comparison with the ones produced by the Robin et al.
(2003) Galactic model, in each direction, for the appropriate field of view,
taking into account the magnitude limit and completeness factor of our
photometry. This test - negative in all cases - is of course truly
significative only for the cases where the field of view was large enough as
to sample even a marginal contaminating population.

The importance of understanding the true origin of outer disk open clusters is
clear. For instance, Be 29, the farthest known OC, is fundamental to study the
metallicity gradient, if its origin is fully Galactic. On the other hand, if
Be 29 is demonstrated to come from another galaxy with a different chemical
history, the interest would be shifted towards understanding which signatures
(if any) characterize the accreted populations, in order to identify and
separate them in the Galactic disk. The Sagittarius dwarf appear to display
chemical peculiarities (Bonifacio et al. 2004, McWilliam \& Smecker-Hane
2005),  and recently Sbordone et al. (2005) claimed that at least one star,
possible member of the CMa, shows chemical signatures  (abundance ratios of
$\alpha$-elements, Cu, and heavy neutron capture elements like La, Ce and Nd)
more appropriate for Local Group dwarf galaxies than for the Milky Way.

Both To 2 and Be 29 have been studied with high resolution spectroscopy; do
they show  similar signatures?    In To 2, Brown et al. (1996) measured
$\alpha$-elements and La. For the latter they find  [La/Fe] = 0.5; for the
former [$\alpha$/Fe] = +0.07 dex, a value in  perfect agreement with many
other old OCs (see e.g. Friel et al. 2003) but unlikely for  dwarf galaxies,
where it tends to be under solar (Venn et al. 2004). Carraro et al. (2004) did
not measure heavy elements in their two Be 29 targets, but  have
$\alpha$-elements, that are said to be enhanced with respect to the solar
values, at variance e.g. with  NGC 2243, which has similar overall metallicity
([Fe/H] around $-$0.4 dex). However, from the values published in their tab.
5, [$\alpha$/Fe] is about 0.08 dex, exactly the average found for NGC 2243 by
Gratton \& Contarini (1994), and again a common value for old open clusters.

\section{Summary}

To summarize the mid-term results of the photometric BOCCE project:
\begin{itemize}
\item We confirm the tight relation between cluster age and magnitude
difference between MS TO and clump luminosity levels, but caution against its
application to young objects, where small number statistics and field
contamination may make the determination of the two luminosity levels too
uncertain.
\item We confirm the anticorrelation between metallicity and distance from the
Galactic center. We don't find evidence of discontinuities around 10-12 kpc
from the center, but our sample is still too small to distinguish between a
smooth gradient and alternative distributions.
\item We find definitely no correlation between cluster age and metallicity.
\item We cannot completely exclude an age-Galactocentric distance relation in
the inner 10-15 kpc.
\end{itemize}

Most of the oldest clusters of the sample (NGC 6253, NGC 6819, Be 29 and Cr
261) show clear evidence of evaporation of their lowest mass stars. Almost all
our clusters appear to have stars (still bound members ?) with the same CMD 
morphology at quite large cluster radii. This circumstance makes it crucial to
always observe control fields at appropriate distances from  the cluster
center.

We emphasize the importance of deriving the cluster parameters with more than
one set of stellar evolution models and with the same photometric conversion
tables to allow for a direct comparison of the different model predictions and
for a correct evaluation of the theoretical uncertainties on the results. We
have not found a set of evolutionary tracks able to reproduce in detail all
the observed features of old and young, metal rich and metal poor clusters.
However, all the three evolution sets adopted in  this project perform quite
well in the comparison with the observational CMDs  and LFs of the examined
clusters, with the BBC models being often found in  better agreement with the
data. Overshooting from convective regions appears to provide a better fit to
the observed properties of stars in the various evolutionary phases, with some
hints that it might be more conspicuous in younger systems. As found also by
Grocholski \& Sarajedini (2003), the CMD  region more difficult to reproduce
is the lower MS, where stellar models often fail to provide the appropriate
shape or slope. We agree with them that the most likely explanation for this
problem is the inadequacy of the adopted photometric conversions, based on
gray atmospheric models, for the lowest mass stars. 

These results are by no means conclusive, since we need to at least double
our  clusters sample to get enough statistics in each age and distance bin.
Moreover, we need to derive accurate chemical abundances from high resolution
spectra for a through interpretation of the apparent metallicity
distributions. Nonetheless, we consider it useful to illustrate the importance
of a homogeneous approach in the derivation of the global cluster properties
and emphasize the need of further accurate studies of these Galactic objects
which can tell us so much more than others on the Milky Way evolution.

\acknowledgements

We thank all the colleagues who have collaborated so far to the photometric
part of the BOCCE project: Gloria Andreuzzi, Lino Bonifazi, Antonio Carusillo,
Luca Di Fabrizio, Flavio     Fusi Pecci, Enrico Gozzoli, Don Hamilton, Gianni
Marconi, Ulisse Munari, Luigi   Pulone, Giuseppina Romeo, Stefano Sandrelli
and Gianni Tessicini. Jason Kalirai has also participated, providing the
beautiful photometric data  from the CFHT.  We also appreciate private
communications on two clusters from Elena Pancino and interesting discussions
with Eileen Friel, who also kindly shared information in advance of
publication. We also wish to thank the anonymous referee for the very
encouraging report.
We have made extensive use of the  BDA, a very useful tool for
which we are  sincerely grateful to J.-C. Mermilliod. This project has been
partially supported by MIUR, under the Cofin project 2003029437.

\clearpage

\begin{deluxetable}{lrrcclllrrcr}
\tablewidth{0pt}
\tabletypesize{\small}
\tablecaption{Cluster parameters derived from the BBC models.
We assumed R$_{\rm GC,\odot}$ = 8 kpc.}
\tablehead{
\colhead{Cluster}           & \colhead{l\degr}      &
\colhead{b\degr}          & \colhead{age}  &
\colhead{\mmm}          & \colhead{\ebv}    &
\colhead{m$_{\rm TO}$ }  & \colhead{Z}  &
\colhead{d} & \colhead{R} & \colhead{$\Delta$V} &
\colhead{ref}\\
\colhead{}           & \colhead{}      &
\colhead{}          & \colhead{ (Gyr)}  &
\colhead{}          & \colhead{}    &
\colhead{(\msun)}  & \colhead{}  &
\colhead{(kpc)} & \colhead{(kpc)} &\colhead{} &
\colhead{}}
\startdata
Cr 261\tablenotemark{a,b}    & 301.68  & $-$5.63  & 6.00: & 12.20: & 0.30:& 1.10: & 0.020 
                                                                   & 2.74 & 6.96  &2.60  & 3 \\
NGC ~2243\tablenotemark{b}   & 239.50  & $-$17.97 & 4.00: & 12.80  & 0.06 & 1.20: & 0.006:
                                                                   & 3.45 & 10.20 &2.20  &2 \\
Be~29     & 197.98 &  8.03   & 3.70 & 15.60 & 0.12 & 1.20  & 0.004 &13.05 & 20.81 & 2.00 &12\\
NGC ~6253 & 335.46 & $-$6.25 & 3.00 & 11.00 & 0.23 & 1.40  & 0.050 & 1.58 &  6.60 & 2.00 &4\\
Be~22     & 199.88 & $-$8.08 & 2.40 & 13.80 & 0.64 & 1.50  & 0.020 & 5.75 & 14.02 & 1.50 &13\\
Be~21     & 186.84 & $-$2.51 & 2.20 & 13.50 & 0.78 & 1.40  & 0.004 & 5.01 & 12.99 & 1.80 &6\\
NGC ~6819 &  73.98 & +8.48   & 2.00 & 12.20 & 0.12 & 1.60  & 0.020 & 2.72 &  7.71 & 2.00 &11\\
NGC ~2506 & 230.56 & +9.94   & 1.70 & 12.60 & 0.05 & 1.70  & 0.020 & 3.26 & 10.38 & 1.60 &5\\
Cr~110    & 209.65 & $-$1.98 & 1.70 & 11.45 & 0.57 & 1.60  & 0.004 & 1.95 &  9.74 & 1.65 &9\\
NGC ~6939 &  95.90 &+12.30   & 1.30 & 11.30 & 0.34 & 1.85  & 0.020 & 1.78 &  8.37 & 1.55 &10\\
Pismis~2  & 258.85 & $-$3.34 & 1.10 & 12.70 & 1.29 & 1.95  & 0.020 & 3.46 &  9.31 & 0.85 &8\\
NGC ~2660 & 265.93 & $-$3.01 & 0.95 & 12.30 & 0.40 & 2.10  & 0.020 & 2.88 &  8.69 & 0.50 &7\\
NGC ~2099 & 177.64 &  3.09   & 0.43 & 10.50 & 0.36 & 2.70  & 0.008 & 1.26 &  9.26 & 0.40 &11\\
NGC ~2168 & 186.59 &  2.22   & 0.18 & ~9.80 & 0.20 & 4.00  & 0.008 & 0.91 &  8.91 & 0.30 &11\\
NGC ~2323 & 221.67 & $-$1.33 & 0.12 & 10.20 & 0.22 & 4.50  & 0.020 & 1.10 &  8.85 & 0.50 &11\\
NGC ~7790 & 116.59 & $-$1.01 & 0.10 & 12.65 & 0.54 & 5.00  & 0.020 & 3.39 &  9.99 & 0.30 &1\\
\enddata
\tablenotetext{a}{Cr 261 is the only cluster of the Table for which the listed
parameters are derived with the FST models rather than with the BBC ones. See
text for details.}
\tablenotetext{b}{Colons indicate more uncertain values, resulting from non
satisfactory  synthetic CMDs} 
\tablerefs{
(1) Romeo et al. 1989; (2) Bonifazi et al. 1990; (3) Gozzoli et al. 1996;
(4) Bragaglia et al. 1997; (5) Marconi et al. 1997: (6) Tosi et al. 1998;
(7) Sandrelli et al. 1999; (8) Di Fabrizio et al. 2001; (9) Bragaglia \& Tosi
2003; (10) Andreuzzi et al. 2004; (11) Kalirai \& Tosi 2004; (12) Tosi et al.
2004; (13) Di Fabrizio et al. 2005.}          
\end{deluxetable}

\begin{deluxetable}{lrrr}
\tablewidth{0pt}
\tablecaption{Forecoming BOCCE clusters; data taken from Dias et al.
(2002). On the first
six work is in progress, while for the others we have only acquired the data. }
\tablehead{
\colhead{Cluster}           & \colhead{l\degr}      &
\colhead{b\degr}     &\colhead{age (Gyr)}     }
\startdata
NGC ~3960  & 294.37  & +6.18	& 1.3 \\ 
Be~17      & 175.65  & $-$3.65  & 12.0\\
Be~32      & 207.95  &  4.40	& 3.4 \\ 
To~2       & 232.83  & $-$6.88  & 1.0 \\ 
Mel~71     & 228.95  & +4.50	& 0.2 \\ 
NGC ~4815  & 303.62  & $-$2.09  & 0.2 \\ 
King~11    & 117.16  &  +06.48  & 1.1 \\ 
NGC ~1193  & 146.81  &  -12.16  & 7.9 \\
NGC ~2204  & 226.01  &  -16.11  & 0.8 \\
NGC ~2477  & 253.56  &  -05.84  & 0.7 \\
NGC ~2849  & 265.27  &  +06.36  & 0.6 \\
NGC ~6134  & 334.92  &  -00.20  & 0.9 \\
NGC ~6603  &  12.86  &  -01.31  & 0.2 \\
Be~19 	   & 176.90  &  -03.59  & 3.1 \\
Be~20 	   & 203.48  &  -17.37  & 6.0 \\
Be~66 	   & 139.43  &  +00.22  & 5.0 \\
IC~361     & 147.48  &  +05.70  & 0.1 \\
IC~1311    &  77.69  &  +04.28  & 0.4 \\
IC~4651    & 340.09  &  -07.91  & 1.1 \\
IC~4756    &  36.38  &  +05.24  & 0.5 \\
Trumpler~5 & 202.86  &  +01.05  & 5.0 \\
\enddata   
\end{deluxetable}
	    
\begin{deluxetable}{lccccccccccc}
\tablewidth{0pt}
\tablecaption{Age, distance modulus and reddening from different stellar models}
\tablehead{
\colhead{Cluster}  &           
\multicolumn{3}{c}{age (Gyr)} & \colhead{} &
\multicolumn{3}{c}{\mmm}      & \colhead{} &
\multicolumn{3}{c}{\ebv} \\
\cline{2-4}  \cline{6-8} \cline{10-12} \\
\colhead{}  &           
\colhead{BBC}  & \colhead{FRA} & \colhead{FST} & \colhead{} &
\colhead{BBC}  & \colhead{FRA} & \colhead{FST} & \colhead{} &
\colhead{BBC}  & \colhead{FRA} & \colhead{FST}}
\startdata
Cr~261    &   9.00 &  5.00 & 6.00 && 12.0 & 12.2 & 12.2 && 0.25 & 0.29 & 0.30 \\
NGC ~2243 &   4.00 &  3.20 & 2.90 && 12.8 & 12.8 & 13.1 && 0.06 & 0.08 & 0.04 \\
Be~29     &   3.70 &  2.30 & 3.50 && 15.6 & 15.8 & 15.8 && 0.13 & 0.15 & 0.10 \\
NGC ~6253 &   3.00 &  3.00 & 3.20 && 11.0 & 10.8 & 10.9 && 0.23 & 0.33 & 0.32 \\
Be~22     &   2.40 &  2.10 & 2.50 && 13.8 & 14.0 & 13.9 && 0.64 & 0.65 & 0.65 \\ 
Be~21     &   2.20 &  2.50 & 2.50 && 13.5 & 13.6 & 13.5 && 0.78 & 0.74 & 0.74 \\
NGC ~6819 &   2.00 &  1.80 & 2.30 && 12.2 & 12.0 & 12.0 && 0.12 & 0.15 & 0.10 \\
NGC ~2506 &   1.70 &  1.60 & 1.80 && 12.6 & 12.6 & 12.7 && 0.00 & 0.07 & 0.04 \\ 
Cr~110    &   1.70 &  1.20 & 1.50 && 11.5 & 11.7 & 11.5 && 0.57 & 0.52 & 0.56 \\
NGC ~6939 &   1.30 &  1.00 & 1.30 && 11.3 & 11.3 & 11.4 && 0.34 & 0.38 & 0.36 \\
Pismis~2  &   1.10 &  1.10 & 1.20 && 12.7 & 12.5 & 12.7 && 1.29 & 1.29 & 1.29 \\
NGC ~2660 &   0.95 &  0.75 & 1.10 && 12.3 & 12.3 & 12.1 && 0.40 & 0.42 & 0.37 \\
NGC ~2099 &   0.43 &  0.59 & 0.52 && 10.5 & 10.5 & 10.4 && 0.34 & 0.16 & 0.36 \\
NGC ~2168 &   0.18 &  0.11 & 0.20 && 9.80 & 9.80 & 9.60 && 0.20 & 0.22 & 0.20 \\
NGC ~2323 &   0.12 &  0.11 & 0.13 && 10.2 & 10.0 & 10.0 && 0.22 & 0.22 & 0.22 \\
NGC ~7790 &   0.10 &  0.05 & 0.10 && 12.7 & 12.7 & 12.7 && 0.54 & 0.54 & 0.54 \\
\enddata
\end{deluxetable}

\clearpage

\begin{deluxetable}{lrrrr}
\tablewidth{0pt}
\tablecaption{Heliocentric radial velocities and metallicities from high
resolution studies for the present sample. References are for both  RV and 
[Fe/H] when the latter is available. We also give for comparison Z, 
the photometric metallicity, taken
from Table 1; Z = 0.004, 0.006, 0.02, 0.05 mean [Fe/H] $\sim$ $-0.7$, $-0.5$,
0.0, +0.4, respectively}
\tablehead{
\colhead{Cluster}  & \colhead{RV}  & \colhead{[Fe/H]} &\colhead{Z}& \colhead{Ref}\\
\colhead{}         &\colhead{kms$^{-1}$} & \colhead{} & \colhead{}& \colhead{}}
\startdata
Cr~261\tablenotemark{a}    & $-$25.9  &  $-$0.03 & 0.020 & 1  \\
NGC ~2243\tablenotemark{a} &   +61.0  &  $-$0.48 & 0.006 & 2  \\
Be~29                      &   +24.6  &  $-$0.44 & 0.004 & 3  \\
NGC ~6253\tablenotemark{a} & $-$26.2  &    +0.36 & 0.050 & 4  \\
Be~22                      &   +95.3  &  $-$0.32 & 0.020 & 5  \\ 
Be~21                      &   +12.4  &  $-$0.54 & 0.004 & 6  \\
NGC ~6819\tablenotemark{a} &    +4.2  &    +0.09 & 0.020 & 7  \\
NGC ~2506\tablenotemark{a} &   +83.7  &  $-$0.20 & 0.020 & 8  \\ 
Cr~110                     &   +40.0  &	         &       & 9       \\
NGC ~6939                  & $-$19.0  &	         &       & 10    \\
Pismis~2                   &   +49.8  &          &	 & 11	 \\
NGC ~2660                  &   +21.2  &          &	 & 12	   \\
NGC ~2099                  &    +7.7  &	         &	 & 13	   \\
NGC ~2168                  &  $-$8.0  &	         &	 & 14	   \\
NGC ~2323                  &    +9.0  &	         &	 & 15	 \\
NGC ~7790                  & $-$78.0  &	         &	 & 16	 \\
\enddata
\tablenotetext{a}{[Fe/H] values measured by our group}	  
\tablerefs{1: Carretta et al. (2005); 2: Gratton \&  Contarini (1994); 3:
Carraro et al. (2004); 4: Carretta et al. (2000);  5: Villanova et al. (2005);
6: Hill \& Pasquini (1999); 7: Bragaglia et al.  (2001); 8: Carretta et al.
(2004); 9:  E. Pancino, priv.comm., preliminary; 10: Milone (1994); 11:
Friel et al. (2002); 12: propr.data, preliminary; 13: Mermilliod et
al. (1996); 14:  Barrado y Navascues et al.  (2001); 15: Harris (1976); 16:
Gorynya et al. (1992)}
\end{deluxetable}

\clearpage

\clearpage

\clearpage

\clearpage

\clearpage

\clearpage

\clearpage

\clearpage

\clearpage

\end{document}